\documentclass[11pt]{article}
\usepackage{xcolor}
\usepackage{graphicx}
\usepackage{epstopdf}
\usepackage[caption=false]{subfig}

\usepackage[numbers,sort&compress]{natbib}
\bibpunct[, ]{[}{]}{,}{n}{,}{,}

\usepackage{amsmath,amssymb,amsthm}
\theoremstyle{plain}

\theoremstyle{definition}

\theoremstyle{remark}

\usepackage{algorithm}
\usepackage{algpseudocode} 
\usepackage{booktabs}

\title{Almost-Exact Simulation Scheme for Heston-type Models: Bermudan and American Option Pricing}

\author{
Mara Kalicanin Dimitrov
\and
Marko Dimitrov
\and
Anatoliy Malyarenko
\and
Ying Ni%
\thanks{Contact: ying.ni@mdu.se}
}

\date{
Division of Mathematics and Physics,\\
M\"alardalen University, V\"aster\aa s, Sweden
}

\begin{document}

\maketitle

\begin{abstract}
Recently, an Almost-Exact Simulation (AES) scheme was introduced for the Heston stochastic volatility model and tested for European option pricing. This paper extends this scheme for pricing Bermudan and American options under both Heston and double Heston models. The AES improves Monte Carlo simulation efficiency by using the non-central chi-square distribution for the variance process. We derive the AES scheme for the double Heston model and compare the performance of the AES schemes under both models with the Euler scheme. Our numerical experiments validate the effectiveness of the AES scheme in providing accurate option prices with reduced computational time, highlighting its robustness for both models. In particular, the AES achieves higher accuracy and computational efficiency when the number of simulation steps matches the exercise dates for Bermudan options. 
\end{abstract}

\noindent\textbf{Keywords:} Almost-Exact Simulation Scheme; Bermudan and American Options; Cox-Ingersoll-Ross Model; Heston Model; Double Heston Model

\section{Introduction}

The challenge of accurate pricing Bermudan and American options, known for their early exercise flexibility, has been a significant area of research in financial mathematics. The underlying price process is often assumed to follow the Heston stochastic volatility model \cite{heston93} (referred to as the Heston model) or sometimes the more complex double Heston stochastic volatility model \cite{christoffersen2009shape} (referred to as the double Heston model). Numerical techniques have been proposed for solving American or Bermudan option problems under Heston model, such as numerically solving partial differential equations (PDEs) \cite{haentjens2015adi, cai2020finite, zvan1998penalty, ikonen2009operator} or using asymptotic expansion method for solving PDEs \cite{zhang2020asymptotic}, Fourier-based valuation method \cite{fang2011fourier}, regression-based methods \cite{longstaff2001valuing} etc.

This study considers the Heston and double Heston models, adopting the widely used regression-based least squares Monte Carlo simulation method for the pricing of Bermudan and American options\cite{longstaff2001valuing}. Bermudan and American option pricing under the Heston model can be efficiently handled using the popular finite difference methods. For the double Heston model with two stochastic volatility processes, the finite difference method can be cumbersome, which motivates studies on an alternative pricing approach using simulation. Moreover, even with the simpler Heston model, a simulation method provides a simple and flexible alternative with several advantages. For example, when the value of the option has a complex path dependency or depends on a basket of several underlying assets, each following Heston-type models, a simulation method is more suited to apply than the finite-difference method.  The focus of this study is, therefore, on simulation schemes for the state variables of the underlying price process specified by stochastic differential equations (SDEs). If a closed-form solution exists for the SDEs, the Monte Carlo simulation can be performed using a bias-free exact simulation scheme; otherwise, the Euler-Maruyama scheme (referred to as the Euler scheme) with discretization bias is often used.

For Bermudan and American option pricing, using an exact simulation scheme instead of the Euler scheme is desirable, as explained in the following simplified example. Consider the Black-Scholes model, where the underlying asset price $S_t$ follows a geometric Brownian motion. The exact simulation scheme for this model is given by the lognormal distribution of $S_t$. To illustrate, a one-year Bermudan option with 5 exercise dates is priced using the least squares Monte-Carlo method. With the exact scheme, the total number of time steps $M$ can be as small as $M = 5$, which is obviously insufficient for the Euler scheme of $S_t$. For an American option, assuming that a Bermudan option with 50 exercise dates per year provides a reasonable approximation (which is often the case in practice), a minimal number of time steps $M = 50$ is sufficient when using the exact scheme. In contrast, using the Euler scheme for $S_t$, a discretization of 50 time steps per year is likely inadequate to achieve a sufficiently small discretization error.

A well-known bias-free exact simulation scheme for the Heston model was introduced in \cite{broadie2006exact}. The authors studied the concept of simulating the sample asset price and variance exactly from their respective distributions, demonstrating that their method exhibits a faster convergence rate than the Euler discretization. However, this exact simulation method is computationally expensive, as highlighted in \cite{andersen2007efficient,lordetal}. To overcome the limitations of this approach, different authors \cite{smith2007almost,andersen2007efficient,chan2010fast,glasserman2011gamma,grzelak2019stochastic} have explored various approximations to the exact simulation schemes.

For example, \cite{smith2007almost} proposed methods that improve the efficiency of simulating stochastic volatility models by reducing the number of small time steps needed to achieve acceptable levels of bias in the simulation error. Their approach, along with others such as those discussed in \cite{chan2010fast,glasserman2011gamma}, involves more sophisticated discretization methods than the Euler scheme. Although these methods reduce the magnitude of the leading error term, they do not change the order of convergence. The innovative exact simulation method by \cite{broadie2006exact} eliminates bias without requiring additional time steps, but it is only effectively applicable to derivative payoffs dependent on observations of the underlying at a few discrete moments of time.

The authors of \cite{chan2010fast} present three novel discretization schemes for the Heston stochastic volatility model. They introduce two schemes for simulating the variance process and one for simulating the integrated variance process, conditional on the initial and end-point of the variance process. Unlike traditional short-time-stepping methods, these new schemes allow the Heston process to evolve accurately over long steps without sampling the intervening values.

In the article \cite{grzelak2019stochastic}, the authors introduced an innovative method for efficiently generating Monte Carlo samples from complex distributions. They presented the Stochastic Collocation Monte Carlo (SCMC) sampler, which operates within a polynomial chaos expansion framework. This technique significantly reduces the computational cost by requiring only a few inversions of the original distribution, supplemented by independent standard normal samples. The authors demonstrated that this approach allows for the efficient and accurate exact simulation of the Heston stochastic volatility model, as originally proposed by \cite{broadie2006exact}.

A simple yet promising approach, known as the Almost Exact Simulation (AES) scheme, is presented in a recent book \cite{oosterlee2019mathematical} for the Heston model. This approach uses the marginal distribution of the CIR process, ensuring that the variance does not become negative by sampling directly from its non-negative distribution. Additionally, one of the two stochastic integrals in the discretized SDE for the underlying asset price process can be rewritten using this exact scheme for the CIR process, making the entire scheme an 'almost' exact scheme. Simulation experiments in \cite{oosterlee2019mathematical} show that AES outperforms the truncated Euler scheme in European call option pricing. 

In this study, the approach by \cite{oosterlee2019mathematical} is chosen due to its elegant simplicity. The experiment in European option pricing implies a conjecture that this AES scheme might show an advantage in American and Bermudan option pricing. To the best of our knowledge, the literature has not yet discussed the use of an exact or almost exact scheme for Bermudan and/or American option pricing. We fill this gap by an extensive investigation into these problems. We first extend the scheme analytically to the double Heston model, then evaluate numerically its accuracy under both the Heston and double Heston models when the number of time steps is minimal. Additionally, the performance of the AES scheme is compared to that of the Euler scheme, particularly in terms of accuracy, computational efficiency, and memory usage. Our numerical simulations demonstrate that the AES scheme significantly enhances simulation accuracy compared to the Euler scheme, mainly when the number of time steps is minimal for in-the-money and at-the-money options.

This article is structured as follows. Section 2 describes Heston and double Heston-type models and pricing methodologies. Section 3 outlines the derivation of the AES schemes. Section 4 presents the results of the numerical experiments, highlighting the accuracy and computational efficiency of the AES schemes in pricing Bermudan and American options under various conditions. Finally, Section 5 concludes the article, summarizing the key contributions and suggesting directions for future research.

\section{Bermudan and American options under Heston-type Models}

Let $(\Omega, \mathcal{F}, \{\mathcal{F}_t\}_{t \in [0,T]}, \mathbb{Q})$ be a complete filtered probability space with a market-given equivalent martingale measure $\mathbb{Q}$. All stochastic processes considered in this study are defined within this probability space, and all expectations are taken under $\mathbb{Q}$. The filtration $\{\mathcal{F}_t\}_{t \in [0,T]}$ represents the history of the market and is assumed to satisfy the usual conditions. This study focuses on the pricing problems of Bermudan and American options with maturity time $T$ on a non-dividend-paying underlying asset whose price at time $t$ is denoted by $S_t$. The underlying price process $(S_t)_{t \in [0,T]}$ follows either a Heston model or a double Heston model. Throughout the paper, a constant risk-free interest rate $r > 0$ is assumed.

The Heston model assumes that the underlying asset price $(S_t)_{t \in [0,T]}$ has a stochastic variance $(\nu_t)_{t \in [0,T]}$ that follows a CIR process, described by the following equations
\begin{align*}\label{eq:Heston}
     dS_t &= rS_t  dt + \sqrt{\nu_t} S_t dW_{t}^{(1)}, \notag  &&  S_0 = s > 0,\\
     d\nu_t &= \kappa \left(\bar{\nu} - \nu_t\right) dt + \gamma \sqrt{\nu_t} dW_{t}^{(2)}, && \nu_0 = v > 0,\\
     d\langle W^{(1)}, W^{(2)} \rangle_t &= \rho_{12} dt, && \rho_{12} \in [-1,1].
\end{align*}
Here, $\kappa > 0$ represents the speed of mean reversion for the variance process, $\bar{\nu} > 0$ refers to the long-term mean, and $\gamma > 0$, known as the \textit{vol-vol}, is the variance coefficient for the process $\nu_t$.

The double Heston model is described by a system of three stochastic differential equations
\begin{align*}
   d S_t &= r S_t dt + \sqrt{\nu_{t}^{(1)}} S_t dW_{t}^{(1)} + \sqrt{\nu_{t}^{(2)}} S_t dW_{t}^{(2)}, \notag  &&  S_0 = s > 0 \\
   d \nu_{t}^{(1)} &= \kappa_{1}\left(\bar{\nu}_{1} - \nu_{t}^{(1)}\right) dt + \gamma_{\nu_{1}} \sqrt{\nu_{t}^{(1)}} dW_{t}^{(3)}, && \nu_{0}^{(1)} = v_1 > 0 \\
  d \nu_{t}^{(2)} &= \kappa_{2}\left(\bar{\nu}_{2} - \nu_{t}^{(2)}\right) dt + \gamma_{\nu_{2}} \sqrt{\nu_{t}^{(2)}} dW_{t}^{(4)}, \notag && \nu_{0}^{(1)} = v_2 > 0 \\
   d\langle W^{(1)}, W^{(3)} \rangle_t &= \rho_{1,3} dt, \quad d\langle W^{(2)}, W^{(4)} \rangle_t = \rho_{2,4} dt, && \rho_{1,3}, \rho_{2,4} \in [-1,1].
\end{align*}
Here, $\kappa_1$ and $\kappa_2$ are the mean reversion rates for the two variance processes, $\bar{\nu}_1$ and $\bar{\nu}_2$ are the long-term mean parameters, and $\gamma_{\nu_{1}}$ and $\gamma_{\nu_{2}}$ are the \textit{vol-vol} parameters. The superscripts 1 and 2 in $\nu_{t}^{(1)}$ and $\nu_{t}^{(2)}$ denote the first and second variance components, respectively, while the superscripts $i$ in $W_t^{(i)},i=1,2,3,4$ identify the correlated Wiener processes driving the dynamics of the asset price and variance components. It is well known that the Feller condition, $2\kappa_i \bar{\nu}_i > \gamma_{\nu_i}^2$ for $i=1,2$, guarantees that the variance processes $\nu_t^{(i)}$ stay positive \cite{feller1951two}. The Feller condition for the Heston model is similar.

Since it is a well-established result that early exercise of American or Bermudan call options on a non-dividend-paying underlying asset is never optimal when the risk-free interest rate is positive, $r>0$, (see, e.g., \cite{hull2016options}), all options considered in this study are put options. We consider a plain vanilla American put option with a strike price $ K $ and maturity time $ T $. To approach the option pricing problem using a simulation method, we start with a time discretization of $[0,T]$. More specifically, the option may be exercised at any time epochs $ 0 < t_1 < \ldots < t_M = T $, equally spaced with step size $\Delta t = T/M$. Let $ h_t = h (S_{t})= (K-S_t)^+$ be the payoff function. If the option is exercised at time epoch $ t_k $, with $ k=1, \ldots, M $, the holder will receive $ h_{t_k} $. At maturity time $ T $, the option's value is given by $ h_T $. At any time prior to $ T $, it is well-known that the American option price is given by
\begin{equation*}
   V_t =  \sup_{\tau \geq t} \mathbb{E}^{\mathbb{Q}}_t \left [e^{-r (\tau-t)} h (S_{\tau}) \right]
\end{equation*}
where $\tau$ is any stopping time taking values in $\{ t_k : k = 1, 2, \ldots, M \} \cap  [t, T]$.

Following the Markov property and the dynamic programming principle, the value of the option at time $ t_k $ is determined as follows:
\begin{align}\label{eq:dpp}
    V_{t_k} = V(t_k, Y_{t_k}) = \max \{h_{t_k}, C ( t_k,  Y_{t_k} )\}
\end{align}
where $ Y_{t_k} := (S_{t_k}, v_{t_k}) $ for the Heston model and $ Y_{t_k} := (S_{t_k}, v^{(1)}_{t_k}, v^{(2)}_{t_k}) $ for the double Heston model. Here, $ C ( t_k,  Y_{t_k} ) $ denotes the unknown continuation function of the option, satisfying
\begin{equation*}
C ( t_k,  Y_{t_k} ) = e^{-r \Delta t} \; \mathbb{E}^{\mathbb{Q}}_{t_k} [ V(t_{k+1},  Y_{t_{k+1}})].
\end{equation*}

To approximate $C(t_k, Y_{t_k})$, this study employs the popular Least Squares Monte Carlo (LSM) method \cite{longstaff2001valuing}, using polynomial regression up to second order in $Y_{t_k}$ with $S_{t_k}$ normalized as $S_{t_k}/K$. The least squares regression is performed using in-the-money paths.

For Bermudan option pricing, the set of exercise dates is only a subset of $\{ t_k : k = 1, 2, \ldots, M \}$. The dynamic programming principle from Eq. \eqref{eq:dpp} should be applied only at this set of exercise dates.

\section{Derivation of AES Schemes}

This section is devoted to showing the derivation of the AES for the double Heston model. To do that, denote $X_t:=\ln S_t$ and let $x_i$  and  $\nu_{i}$ be the discrete approximations to $X_{t_i}$ and $ \nu_{t_i}$ respectively for $ i = 0, 1, \ldots M$ ,  the AES for the Heston model is given by \cite{oosterlee2019mathematical} as follows
\begin{align*}
  x_{i+1} \approx  x_{i}+c_0+c_1 \nu_{i}+c_2 \nu_{i+1}+\sqrt{c_3 \cdot \nu_{i}}Z_1,  
\end{align*}
where $Z_1$ is drawn from the standard normal distribution and the CIR process \cite{cox1985theory} simulated as follows
\begin{align}\label{eq:HestonAESVariance}
    \nu_{i+1}\approx \bar{c}(t_{i+1}, t_{i})\cdot \chi^{2}\left(\delta, \bar{\kappa}\left(t_{i+1}, t_{i}\right)\right),
\end{align}
with the following parameters:
\begin{align*}
\bar{c}\left(t_{i+1}, t_{i}\right) &=\frac{\gamma^2}{4 \kappa}\left(1-\mathrm{e}^{-\kappa\left(t_{i+1}-t_{i}\right)}\right), \quad 
\bar{\kappa}\left(t_{i+1}, t_{i}\right) =\frac{4 \kappa \mathrm{e}^{-\kappa\left(t_{i+1}-t_{i}\right)}}{\gamma^{2}\left(1-\mathrm{e}^{-\kappa\left(t_{i+1}-t_{i}\right)}\right)} \cdot \nu_{i}, 
\end{align*}
and $\chi^2(\delta,\bar{k}(\cdot , \cdot))$ is sampled from the noncentral chi-squared distribution with $\delta$ degrees of freedom and noncentrality parameter $\bar{k}(\cdot , \cdot)$. The parameters are 
$$
\begin{aligned}
c_0&=\left(r- \frac{\rho_{1, 2}}{\gamma}\kappa \bar{\nu}\right)\Delta t, 
 c_1=\left(\frac{\rho_{1, 2}}{\gamma}\kappa-\frac{1}{2} \right)\Delta t -\frac{\rho_{1, 2}}{\gamma},  
 c_2=\frac{\rho_{1, 2}}{\gamma}, 
 c_3=\left(1-\rho_{1,2}^{2} \right)\Delta t.
\end{aligned}
$$

Let us extend this idea of AES to the double Heston model. By applying It\^{o}'s lemma to the log transformation $X_t:=\ln S_t$ and the Cholesky decomposition of the correlation matrix 
\begin{align*}
    dW_{t}^{(1)} &= \rho_{1,3} d\widetilde{W}_t^{(3)} + \sqrt{1 - \rho_{1,3}^2} d\widetilde{W}_t^{(1)}, \\
    dW_{t}^{(2)} &= \rho_{2,4} d\widetilde{W}_t^{(4)} + \sqrt{1 - \rho_{2,4}^2} d\widetilde{W}_t^{(2)}, \\
    dW_{t}^{(3)} &= d\widetilde{W}_t^{(3)}, \\
    dW_{t}^{(4)} &= d\widetilde{W}_t^{(4)},
\end{align*}
the following holds
\begin{gather*}
    \begin{split}
        d X_t&=\left(r-\frac{1}{2}\left(\nu_{t}^{(1)}+\nu_{t}^{(2)} \right) \right)d t+\sqrt{\nu_{t}^{(1)}} \left(\rho_{1,3} d\widetilde{W}_{t}^{(3)}+\sqrt{1-\rho^2_{1,3}} d\widetilde{W}_{t}^{(1)}\right)  \\
&\phantom{=} +\sqrt{\nu_{t}^{(2)}} \left(\rho_{2,4}  d\widetilde{W}_{t}^{(4)}+\sqrt{1-\rho^2_{2,4}} d\widetilde{W}_{t}^{(2)}\right),\\
d \nu_{t}^{(1)}&=\kappa_{1}\left(\bar{\nu}_{1}-\nu_{t}^{(1)}\right) d t+\gamma_{v_{1}} \sqrt{\nu_{t}^{(1)}} d \widetilde{W}_{t}^{(3)},\\
d \nu_{t}^{(2)}&=\kappa_{2}\left(\bar{\nu}_{2}-\nu_{t}^{(2)}\right)d t+\gamma_{v_{2}} \sqrt{\nu_{t}^{(2)}} d \widetilde{W}_{t}^{(4)},
    \end{split}
\end{gather*}
where $d \widetilde{W}_{t}^{(1)},d \widetilde{W}_{t}^{(2)},d \widetilde{W}_{t}^{(3)}$ and $d \widetilde{W}_{t}^{(4)}$ are independent Brownian increments. Note that the Cholesky decomposition is applied in this way so variance components are driven by independent Brownian motions in order to use properties of marginal distribution of the variance processes.

This approach results in a system of three SDEs, where each variance process is driven independently by Brownian motions, and $X_t$ is correlated with $\nu_t^{(i)}$ for $i=1,2$. Consequently, the properties of the marginal distribution of $\nu_t^{(i)}$ can be used.

Taking integration of all processes in a certain time interval $[t_i,t_{i+1}],$ the following discretization scheme is obtained
\begin{align*} 
x_{i+1} &= x_{i} + \int_{t_{i}}^{t_{i+1}} \left(r - \frac{1}{2} \left(\nu_{t}^{(1)} + \nu_{t}^{(2)}\right)\right) dt + \rho_{1,3} \int_{t_{i}}^{t_{i+1}} \sqrt{\nu_{t}^{(1)}} d \widetilde{W}_t^{(3)}  \\
&\phantom{= x_{i2}} + \sqrt{1 - \rho_{1,3}^{2}} \int_{t_{i}}^{t_{i+1}} \sqrt{\nu_{t}^{(1)}} d \widetilde{W}_t^{(1)} + \rho_{2,4} \int_{t_{i}}^{t_{i+1}} \sqrt{\nu_{t}^{(2)}} d \widetilde{W}_t^{(4)}  \\
&\phantom{= x_{i2}} + \sqrt{1 - \rho_{2,4}^{2}} \int_{t_{i}}^{t_{i+1}} \sqrt{\nu_{t}^{(2)}} d \widetilde{W}_t^{(2)}, \\ 
\nu_{i+1}^{(1)} &= \nu_{i}^{(1)} + \kappa_1 \int_{t_{i}}^{t_{i+1}} \left(\bar{\nu}_{1} - \nu_{t}^{(1)}\right) dt + \gamma_{\nu_{1}} \int_{t_{i}}^{t_{i+1}} \sqrt{\nu_{t}^{(1)}} d \widetilde{W}_t^{(3)},  \\
\nu_{i+1}^{(2)} &= \nu_{i}^{(2)} + \kappa_2 \int_{t_{i}}^{t_{i+1}} \left(\bar{\nu}_{2} - \nu_{t}^{(2)}\right) dt + \gamma_{\nu_{2}} \int_{t_{i}}^{t_{i+1}} \sqrt{\nu_{t}^{(2)}} d \widetilde{W}_t^{(4)}. 
\end{align*}

It is important to note that the same integrals involving $\widetilde{W}_t^{(i)}$ for $i=3,4$ appear in the SDEs. Using this, the discretization for $x_{i+1}$ is given by:
\begin{align*}
x_{i+1} &= x_{i} + \int_{t_{i}}^{t_{i+1}} \left(r - \frac{1}{2} \left(\nu_{t}^{(1)} + \nu_{t}^{(2)}\right) \right) dt + \sqrt{1 - \rho_{1,3}^{2}} \int_{t_{i}}^{t_{i+1}} \sqrt{\nu_{t}^{(1)}} d \widetilde{W}_t^{(1)} \\
&\phantom{= x_{i}} + \frac{\rho_{1,3}}{\gamma_{\nu_{1}}} \left(\nu_{i+1}^{(1)} - \nu_{i}^{(1)} - \kappa_1 \int_{t_{i}}^{t_{i+1}} \left(\bar{\nu}_{1} - \nu_{t}^{(1)}\right) dt\right) \\
&\phantom{= x_{i}} + \frac{\rho_{2,4}}{\gamma_{\nu_{2}}} \left(\nu_{i+1}^{(2)} - \nu_{i}^{(2)} - \kappa_2 \int_{t_{i}}^{t_{i+1}} \left(\bar{\nu}_{2} - \nu_{t}^{(2)}\right) dt\right) \\
&\phantom{= x_{i}} + \sqrt{1 - \rho_{2,4}^{2}} \int_{t_{i}}^{t_{i+1}} \sqrt{\nu_{t}^{(2)}} d \widetilde{W}_t^{(2)}.
\end{align*}
As in the Euler scheme, the remaining two stochastic integrals involving $\widetilde{W}_t^{(i)}, i=1,2$ are approximated by their left-side integration boundary values of the integrand. Using this approach, since $\Delta t = t_{i+1} - t_i$, the discretization for $x_{i+1}$ is given by:
\begin{align}\label{eq:DH_diskt}
x_{i+1} &\approx x_{i} + \left(r - \frac{1}{2} \left(\nu_{i}^{(1)} + \nu_{i}^{(2)} \right)\right) \Delta t 
+ \frac{\rho_{1,3}}{\gamma_{\nu_{1}}} \left(\nu_{i+1}^{(1)} - \nu_{i}^{(1)} - \kappa_1 \left(\bar{\nu}_{1} - \nu_{i}^{(1)}\right) \Delta t\right) \notag \\
&\phantom{=} + \sqrt{1 - \rho_{1,3}^{2}} \sqrt{\nu_{i}^{(1)}} \left(\widetilde{W}_{t_{i+1}}^{(1)} - \widetilde{W}_{t_{i}}^{(1)}\right)
+ \frac{\rho_{2,4}}{\gamma_{\nu_{2}}} \left(\nu_{i+1}^{(2)} - \nu_{i}^{(2)} - \kappa_2 \left(\bar{\nu}_{2} - \nu_{i}^{(2)}\right) \Delta t\right) \notag \\
&\phantom{=} + \sqrt{1 - \rho_{2,4}^{2}} \sqrt{\nu_{i}^{(2)}} \left(\widetilde{W}_{t_{i+1}}^{(2)} - \widetilde{W}_{t_{i}}^{(2)}\right).
\end{align}
By using the properties of normally distributed random variables and substituting into Eq. \eqref{eq:DH_diskt}, the AES scheme for the double Heston model is given by
\begin{align*}
    x_{i+1} \approx x_{i} + c_0 + c_1 \nu_{i}^{(1)} + c_2 \nu_{i}^{(2)} + c_3 \nu_{i+1}^{(1)} + c_4 \nu_{i+1}^{(2)} +
    \sqrt{c_5 \cdot \nu_{i}^{(1)}} Z_1 + \sqrt{c_6 \cdot \nu_{i}^{(2)}} Z_2,
\end{align*}
where $Z_1, Z_2$ are independent draws from the standard normal distribution, and the two CIR processes are
\begin{align*}
    \nu_{i+1}^{(1)} &\approx \bar{c}^{(1)}(t_{i+1}, t_{i}) \cdot \chi^{2} \left(\delta^{(1)}, \bar{\kappa}^{(1)} \left(t_{i+1}, t_{i}\right)\right), \\
    \nu_{i+1}^{(2)} &\approx \bar{c}^{(2)}(t_{i+1}, t_{i}) \cdot \chi^{2} \left(\delta^{(2)}, \bar{\kappa}^{(2)} \left(t_{i+1}, t_{i}\right)\right).
\end{align*}
The parameters are given by:
\begin{align*}
\bar{c}^{(j)}\left(t_{i+1}, t_{i}\right) &= \frac{\gamma_{\nu_{j}}^{2}}{4 \kappa_{j}} \left(1 - \mathrm{e}^{-\kappa_{j}\left(t_{i+1} - t_{i}\right)}\right), &&
\bar{\kappa}^{(j)}\left(t_{i+1}, t_{i}\right) = \frac{4 \kappa_{j} \mathrm{e}^{-\kappa_{j}\left(t_{i+1} - t_{i}\right)} \cdot \nu_{i}^{(j)}}{\gamma_{\nu_{j}}^{2} \left(1 - \mathrm{e}^{-\kappa_{j}\left(t_{i+1} - t_{i}\right)}\right)},
\end{align*}
for $j = 1, 2$. The constants are:
$$
\begin{aligned}
c_0 &= \left(r - \frac{\rho_{1,3}}{\gamma_{\nu_{1}}} \kappa_1 \bar{\nu}_{1} - \frac{\rho_{2,4}}{\gamma_{\nu_{2}}} \kappa_2 \bar{\nu}_{2}\right) \Delta t, &&
c_3 = \frac{\rho_{1,3}}{\gamma_{\nu_{1}}}, &&& 
c_5 = \left(1 - \rho_{1,3}^{2}\right) \Delta t, \\
c_1 &= \left(\frac{\rho_{1,3}}{\gamma_{\nu_{1}}} \kappa_1 - \frac{1}{2}\right) \Delta t - \frac{\rho_{1,3}}{\gamma_{\nu_{1}}}, &&
c_4 = \frac{\rho_{2,4}}{\gamma_{\nu_{2}}}, &&& 
c_6 = \left(1 - \rho_{2,4}^{2}\right) \Delta t, \\
c_2 &= \left(\frac{\rho_{2,4}}{\gamma_{\nu_{2}}} \kappa_2 - \frac{1}{2}\right) \Delta t - \frac{\rho_{2,4}}{\gamma_{\nu_{2}}}.
\end{aligned}
$$

To close this section, a brief comparison of the AES schemes and the more conventional Euler scheme is presented. Taking the double Heston model as an example, a significant advantage of the AES scheme is the complete removal of the well-known discretization bias inherent in truncating or reflecting Euler schemes for the CIR variance processes (see \cite{lordetal} for a discussion on the discretization bias). Additionally, two out of the four stochastic integrals in the discretization scheme for the log-price $x_{i+1}$, specifically 
$$
\int_{t_{i}}^{t_{i+1}} \sqrt{\nu_{t}^{(1)}} d \widetilde{W}_t^{(3)} \text{ and } \int_{t_{i}}^{t_{i+1}} \sqrt{\nu_{t}^{(2)}} d \widetilde{W}_t^{(4)},$$ 
are replaced by more accurate expressions in terms of the exactly simulated variance realizations. While the resulting simulation scheme is not fully 'exact' due to the remaining two stochastic integrals for the log-price being approximated by the Euler scheme, it is considered 'almost' exact. Consequently, the discretization error is expected to be very small compared to the Euler counterpart. The next section conducts extensive experimental studies to evaluate how well the AES schemes approximate an exact simulation scheme.

\section{Evaluation of AES Schemes}
This section focuses on numerical tests of the AES schemes presented in the previous section, compared to the Euler scheme where the simulated variances $\nu_{i}$ are replaced by their positive approximations. The non-central chi-square distribution is sampled using the function \texttt{noncentral\_chisquare} from Python \textit{numpy.random} library.  The examples considered are Bermudan and American put option pricing problems under the Heston and double Heston models. For each option pricing problem, 20 simulation runs are conducted, and the average price is reported as the estimated Monte Carlo price. By observation, the variation in results across the 20 simulation runs is consistently low for both AES and Euler schemes, indicating robust performance. The reported average prices across multiple runs, coupled with the low variability, provide strong evidence of accuracy and reliability without the need for additional error metrics. The presented computational experiments were conducted using Python on a laptop with an 11th Gen Intel(R) Core(TM) i7-11800H CPU @ 2.3GHz, 32 GB RAM, and an NVIDIA T1200 GPU.

The state process paths are simulated according to the time grid of $0 < t_1 < \ldots < t_M = T$. For Bermudan option pricing, a set of equidistant exercise points is considered, each coinciding with a time point in the time grid. The conjecture is that for a decent price approximation using the AES, $M$ can be set equal to the total number of exercise dates on $[0,T]$; that is, the simulation steps are exactly the exercise points. For American option pricing, this price is approximated by a Bermudan option with at least 48 exercise points per year, implying $M=12$ for $T=0.25$. While a higher value of $M$ will yield a more accurate approximation due to higher exercising frequency, it is insightful to examine the accuracy of the AES with such a small value of $M=12$ to demonstrate its 'almost exactness'.

The implementation details of all simulation schemes are given in Appendix A. In particular,  Algorithms \ref{alg:AES} and \ref{alg:DoubleHestonAES} illustrate the implementation steps of the AES scheme for the Heston and Double Heston models.

Furthermore, the pricing of American and Bermudan options was performed using the LSM method \cite{longstaff2001valuing}, applying polynomial regression to in-the-money paths by constructing features that included the normalized asset prices ($S/K$), their squares, the variance components, their squares, and interaction terms, with a constant term added; these features were used to estimate continuation values and determine optimal exercise decisions by comparing them with immediate exercise values.

\subsection{AES under the Heston model}

Two sets of parameters for the Heston model are considered: one widely used satisfies the Feller condition and has a positive correlation coefficient, and another is more realistic with a negative correlation but does not satisfy the Feller condition.

 The first set, employed in various articles including \cite{clarke1996multigrid,haentjens2015adi,ikonen2009operator,oosterlee2003multigrid,persson2010pricing,vellekoop2009tree,zvan1998penalty}, is given by
\begin{equation}
\begin{aligned}
K &= 10, &\quad v_0 &= 0.0625, &\quad T &= 0.25, &\quad r &= 0.1, \\
\gamma &= 0.9, &\quad \bar{\nu} &= 0.16, &\quad \kappa &= 5, &\quad \rho_{1,2} &= 0.1.
\end{aligned}
\label{eq:ParametersFellerConditionHolds}
\end{equation}
This set follows the Feller condition and has a positive correlation coefficient. The second set of parameters considered is
\begin{equation}
\begin{aligned}
K &= 100, &\quad v_0 &= 0.0348, &\quad T &= 0.25, &\quad r &= 0.04, \\
\gamma &= 0.39, &\quad \bar{\nu} &= 0.0348, &\quad \kappa &= 1.15, &\quad \rho_{1,2} &= -0.64.
\end{aligned}
\label{eq:ParametersFellerConditionDoesNotHold}
\end{equation}
Notice that this set of parameters does not follow the Feller condition. Article \cite{haentjens2015adi} conducted a study on this particular parameter set for the pricing of American put options while \cite{fang2011fourier} used them for Bermudan option pricing.

Throughout the numerical analysis for Bermudan options, the simulation for all experiments will consistently have a fixed number of paths, with $N = 1,000,000$, and a total of $20$ runs will be conducted for each experiment.

For illustration, the Algorithm \ref{alg:AES} in Appendix A outlines the steps involved in implementing the AES scheme for simulating paths of the Heston model.

\textbf{Bermudan Options Experiment 1.} This experiment investigates the accuracy of the AES scheme in pricing Bermudan options and compares it to the Euler scheme. The numerical analysis uses the parameters provided in Eq. \eqref{eq:ParametersFellerConditionDoesNotHold} and reference prices obtained from \cite{fang2011fourier}. The options are tested for 20, 40, and 60 exercise dates, corresponding to minimum time steps \( M = 20, 40, 60 \), respectively.

The results for 20 exercise dates, where \( M = 20 \) for AES and \( M = 40 \) for Euler to provide a fair accuracy comparison, are presented in Table \ref{table:BermudanNonFeller}. It is observed that the AES scheme achieves prices close to the reference values, demonstrating excellent accuracy even with the minimal number of steps. This is particularly evident for in-the-money (\( S_0 = 90 \)) and at-the-money (\( S_0 = 100 \)) options, where the AES scheme outperforms the Euler scheme in both accuracy and computational efficiency. For out-of-the-money options (\( S_0 = 110 \)), the accuracy of AES remains acceptable but slightly lower.

\begin{table}[htbp]
\centering
\caption{Prices of Bermudan options under the Heston model using AES and Euler schemes, including average computation time in seconds.}
\label{table:BermudanNonFeller}
\begin{tabular}{ccccc}
\toprule
Scheme & No. of Steps & $S_0 = 90$ & $S_0 = 100$ & $S_0 = 110$ \\
\midrule
AES    & 20 & 9.966 & 3.195 & 0.917 \\
Time (s) &  & (12.35) & (8.78) & (6.04) \\
\midrule
Euler  & 40 & 9.956 & 3.189 & 0.924 \\
Time (s) &  & (15.73) & (12.82) & (9.60) \\
\midrule
Reference Price &  & 9.978 & 3.205 & 0.927 \\
\bottomrule
\end{tabular}
\end{table}

A comparison of the AES and Euler schemes highlights the efficiency of the AES scheme. To achieve accuracy comparable to AES, the Euler scheme requires approximately double the number of steps, resulting in increased computational time and memory usage. This trade-off becomes particularly significant for simulations requiring fine time discretization, as the Euler scheme's memory requirements scale significantly with the number of steps.

\begin{figure}
\centering
\subfloat[Relative Errors for Bermudan options with $40$ exercise dates.]{%
\resizebox*{6.8cm}{!}{\includegraphics{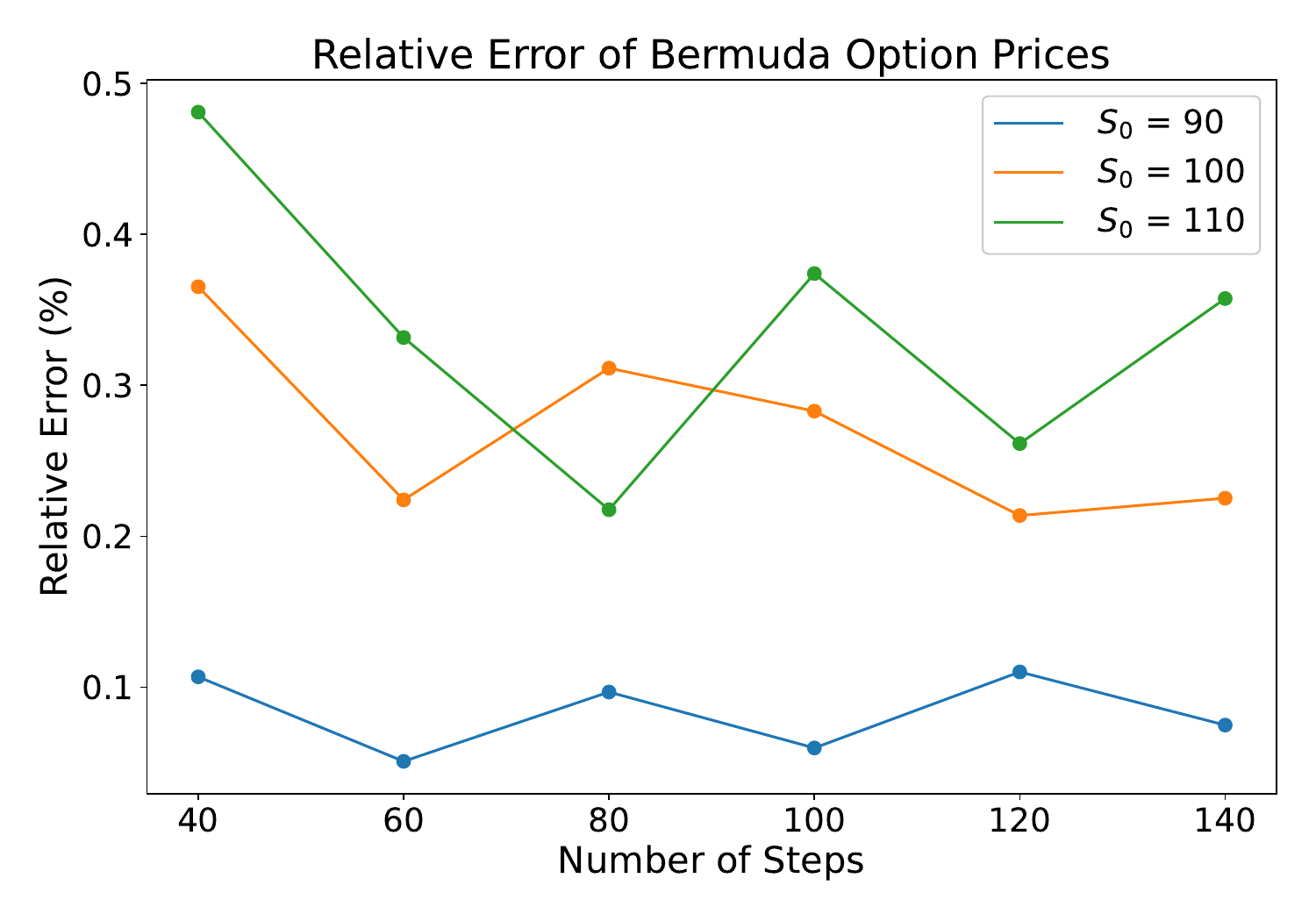}}}\hspace{10pt}
\subfloat[Relative Errors for Bermudan options with $60$ exercise dates.]{%
\resizebox*{6.8cm}{!}{\includegraphics{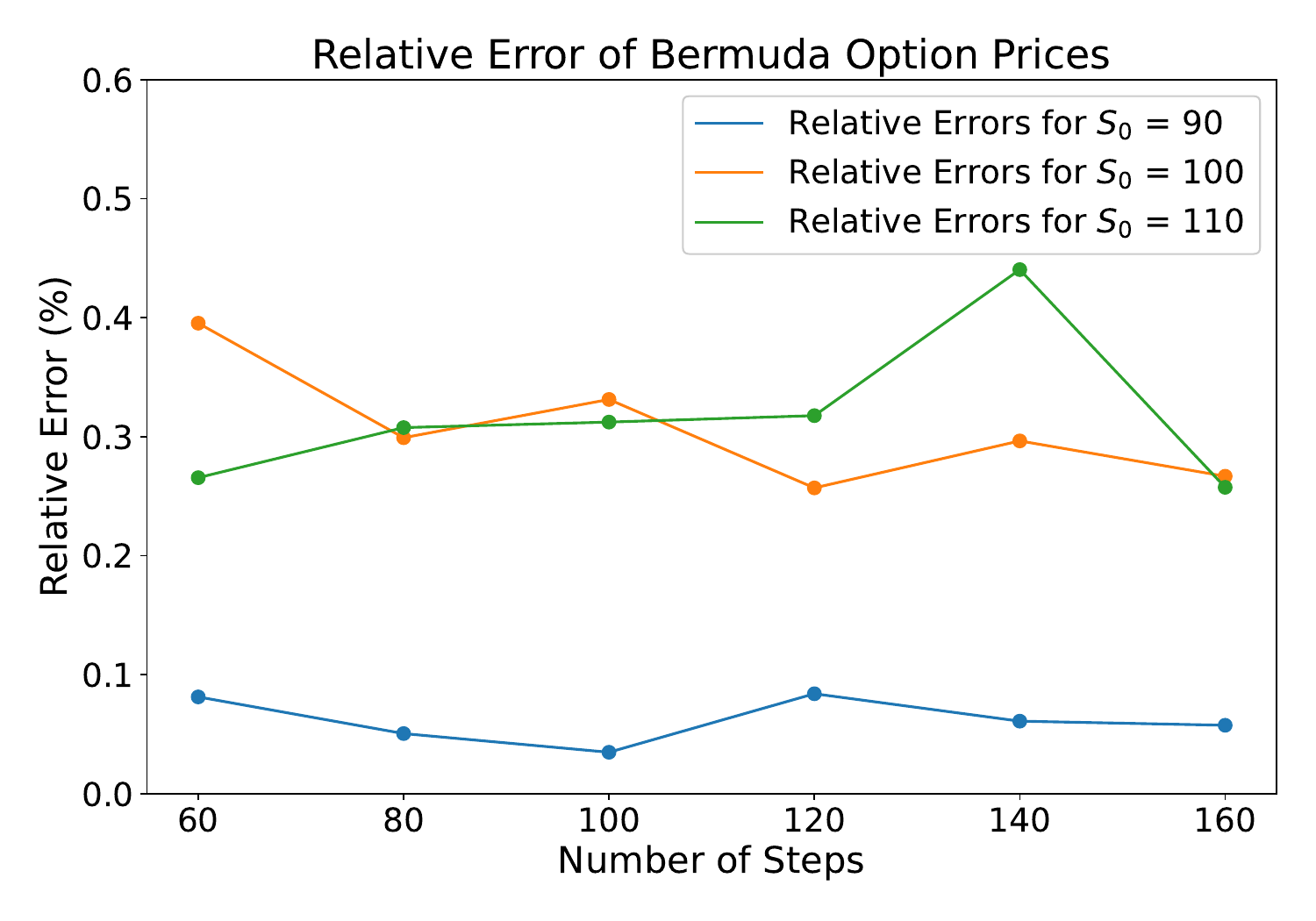}}}
\caption{Relative Errors of Bermudan option prices with parameters given in Eq. \eqref{eq:ParametersFellerConditionDoesNotHold} calculated using AES under the Heston Model. The results highlight the accuracy of the AES scheme across different initial asset prices and exercise dates.}
\label{fig:RelativeErrors_BermudanOptions}
\end{figure}

The behavior of relative errors for 40 and 60 exercise dates is illustrated in Fig. \ref{fig:RelativeErrors_BermudanOptions}. 
\begin{itemize}
    \item Figure \ref{fig:RelativeErrors_BermudanOptions}(a): For 40 exercise dates, the AES scheme achieves low relative errors, especially for in-the-money options (\( S_0 = 90 \)), where the error is below $0.1\%$. The relative errors for at-the-money (\( S_0 = 100 \)) and out-of-the-money (\( S_0 = 110 \)) options remain within $0.2–0.4\%$, reflecting the stability of the AES.
    
    \item Figure \ref{fig:RelativeErrors_BermudanOptions}(b): For 60 exercise dates, the trend in relative errors remains consistent with the results for 40 exercise dates. 
\end{itemize}

These findings show that the AES scheme achieves high accuracy even with a minimal number of time steps, making it a computationally efficient alternative to the Euler scheme. This is particularly advantageous for Bermudan options, where reducing time steps directly reduces computational cost without sacrificing accuracy. 

\textbf{Bermudan Options Experiment 2.} In this experiment, we compute bi-weekly Bermudan option prices under the Heston model and compare the accuracy of the AES scheme to the Euler scheme. Since no established reference prices exist for Bermudan options across varying times to maturity, we generate reference prices using the Euler scheme with \( N = 1,000,000 \) paths and \( M = 750 \) steps to ensure accuracy.

\begin{figure}
\centering
\subfloat[Relative Errors when $S_0 = 90$.]{%
\resizebox*{7cm}{!}{\includegraphics{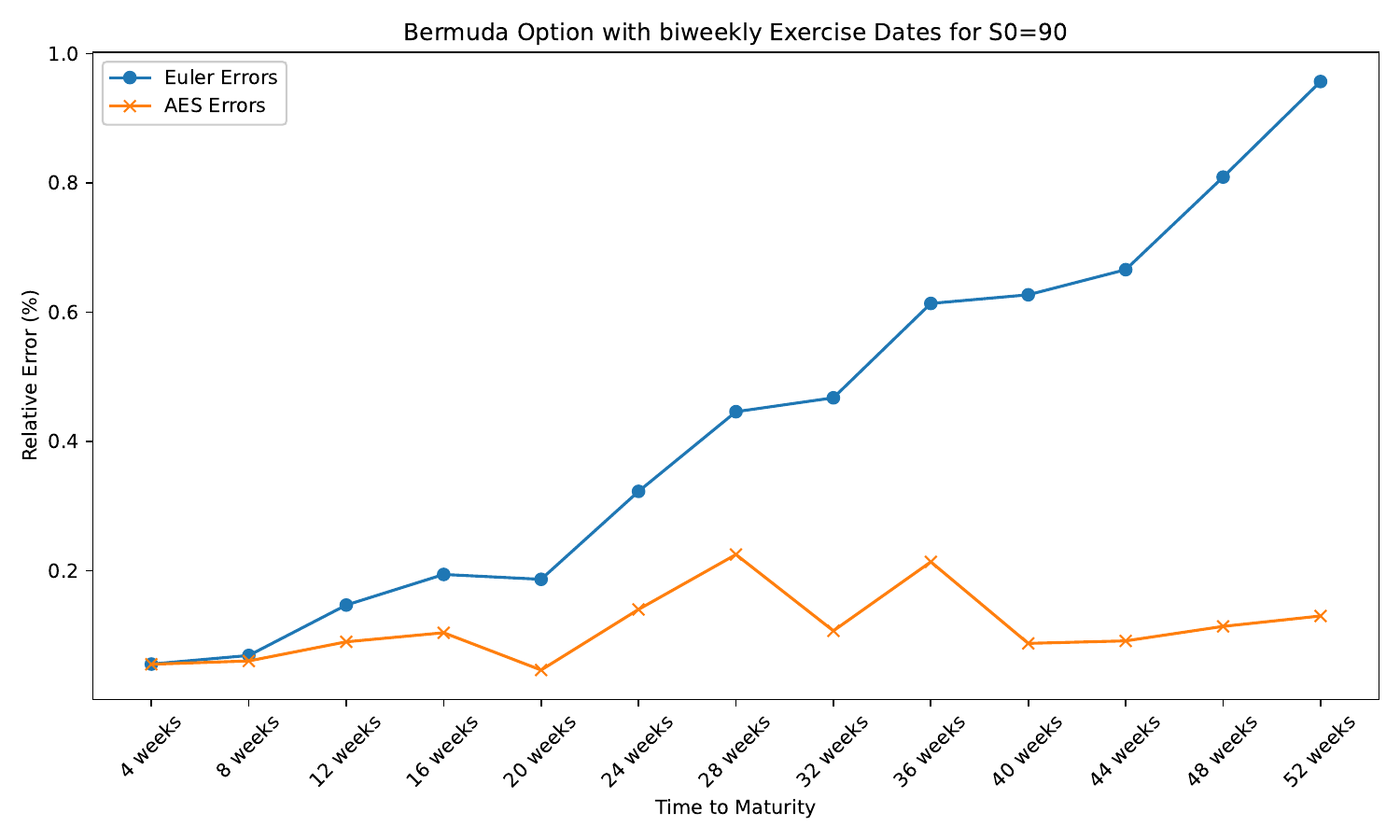}}}\hspace{5pt}
\subfloat[Relative Errors when $S_0 = 100$.]{%
\resizebox*{7cm}{!}{\includegraphics{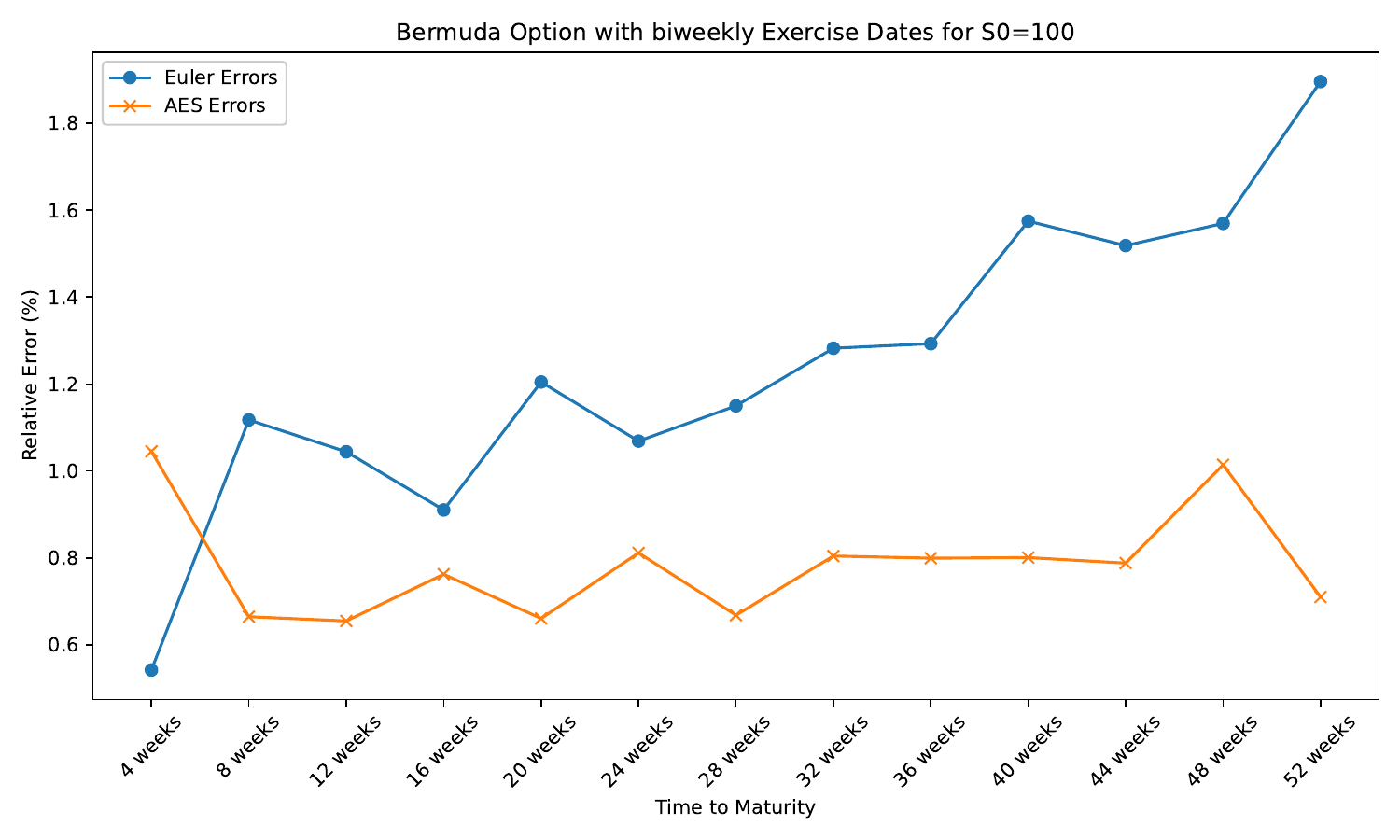}}}\hspace{5pt}
\subfloat[Relative Errors when $S_0 = 110$.]{%
\resizebox*{7cm}{!}{\includegraphics{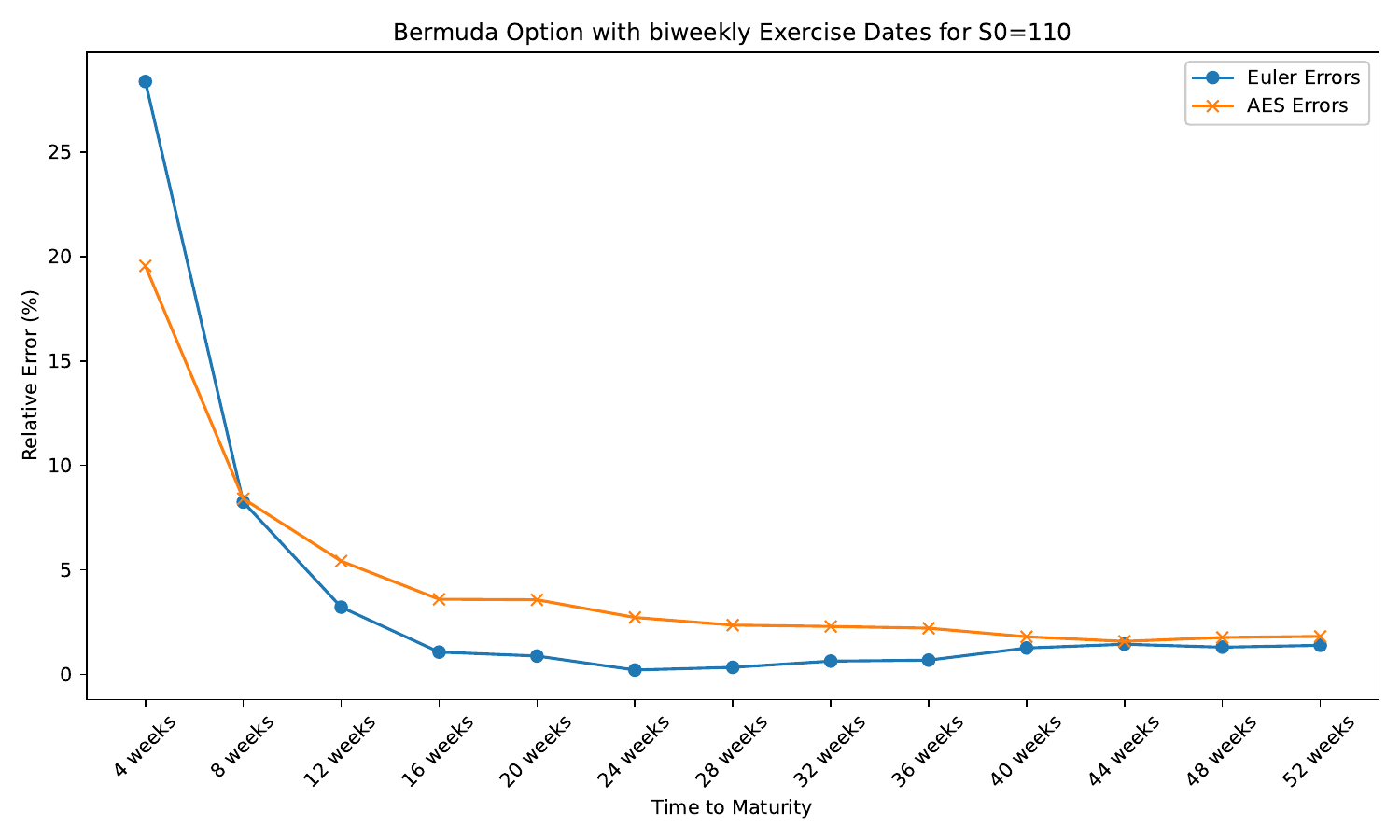}}}
\caption{Relative Errors of biweekly Bermudan option values with parameters given in Eq. \ref{eq:ParametersFellerConditionDoesNotHold} calculated under the Heston Model using AES and Euler schemes for different initial asset prices $S_0$. The AES scheme consistently achieves lower errors for in-the-money and at-the-money options, especially as the time to maturity increases.}
\label{fig:RelativeErrors_BermudanOptions_S0}
\end{figure}

The number of steps is set equal to the number of exercise dates, ranging from 2 exercise dates (4 weeks) up to 26 exercise dates (52 weeks). The relative errors for different initial asset prices (\( S_0 = 90, 100, 110 \)) are illustrated in Fig. \ref{fig:RelativeErrors_BermudanOptions_S0}.  

\begin{itemize}
    \item Figure \ref{fig:RelativeErrors_BermudanOptions_S0}(a): For \( S_0 = 90 \) (in-the-money options), the AES scheme consistently achieves lower relative errors than the Euler scheme across all maturities. The relative error for AES remains stable and below $0.3\%$, while the Euler scheme's error increases significantly with maturity.

    \item Figure \ref{fig:RelativeErrors_BermudanOptions_S0}(b): For \( S_0 = 100 \) (at-the-money options), the AES scheme outperforms the Euler scheme. As the maturity increases, the AES consistently maintains a clear accuracy advantage.

    \item Figure \ref{fig:RelativeErrors_BermudanOptions_S0}(c): For \( S_0 = 110 \) (out-of-the-money options), the relative errors for both AES and Euler schemes are higher at short maturities. However, as time to maturity increases, the errors for both schemes decrease and converge. Notably, the AES scheme remains competitive with the Euler scheme, demonstrating similar accuracy for long maturities.
\end{itemize}

These findings confirm the robustness and accuracy of the AES scheme.

\begin{figure}
\centering
\subfloat[Accuracy difference between AES and Euler schemes.]{%
\resizebox*{7cm}{!}{\includegraphics{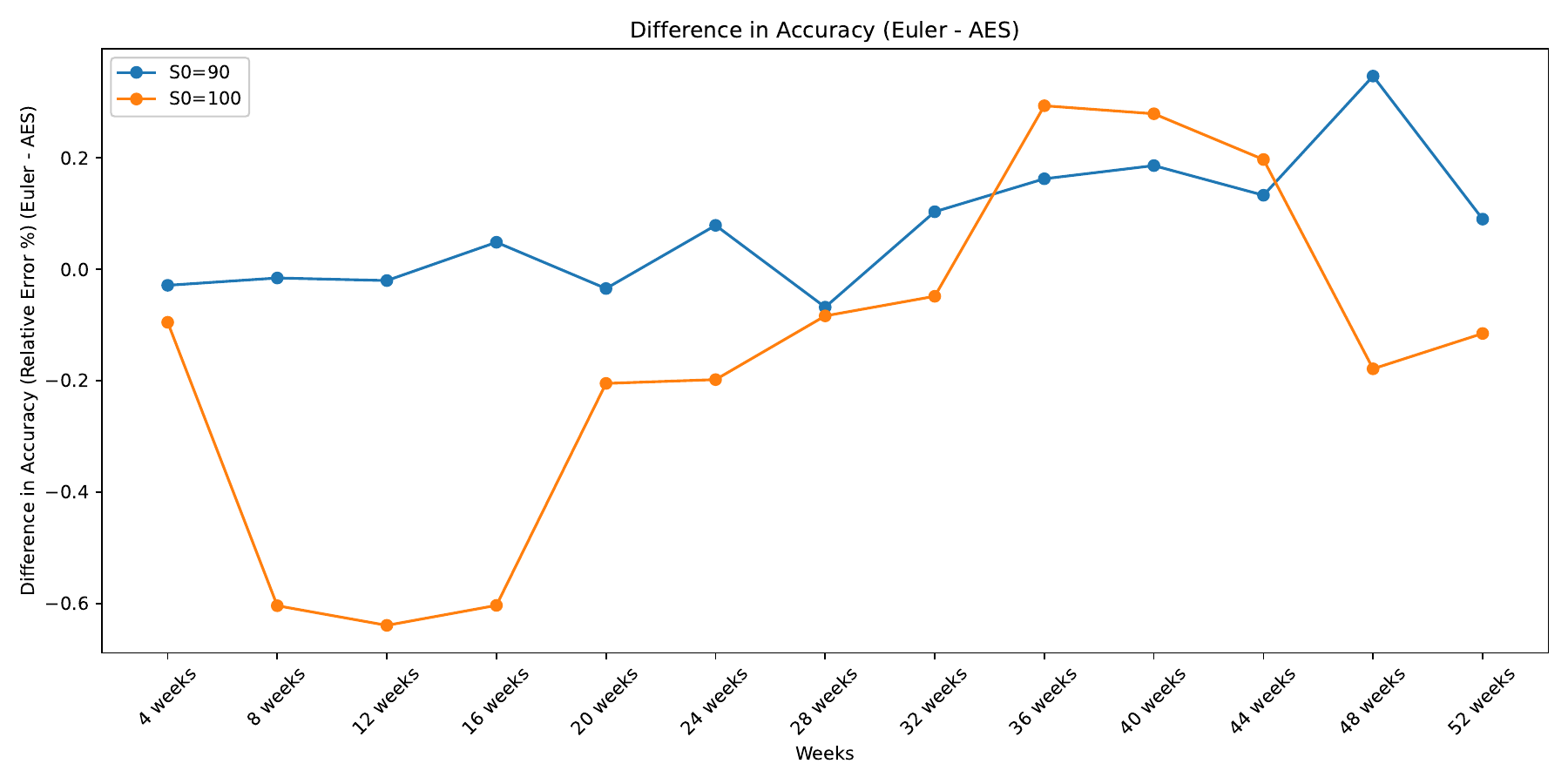}}}\hspace{5pt}
\subfloat[Computation time difference between AES and Euler schemes.]{%
\resizebox*{7cm}{!}{\includegraphics{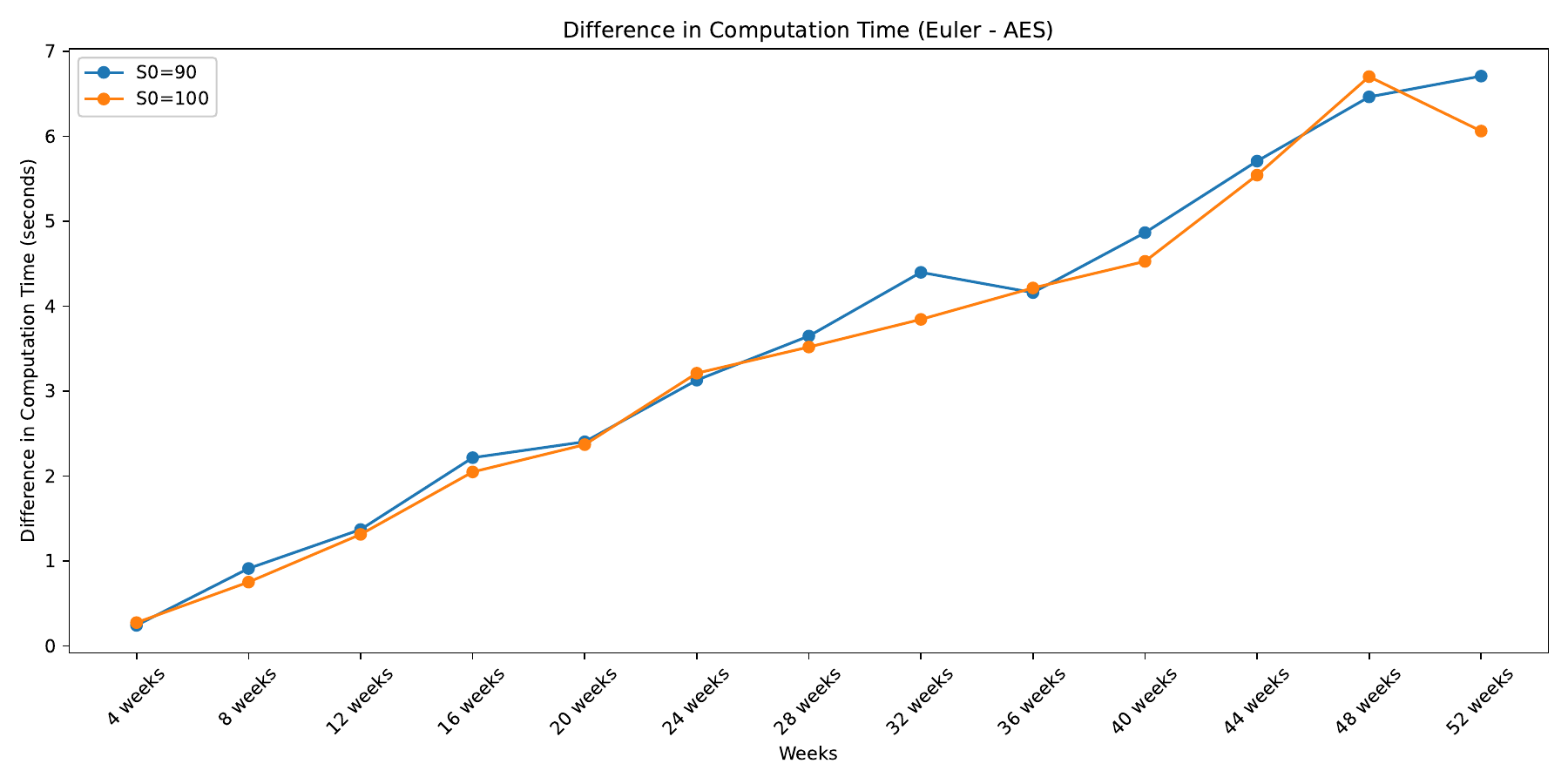}}}\hspace{5pt}
\subfloat[Memory usage difference between AES and Euler schemes.]{%
\resizebox*{7cm}{!}{\includegraphics{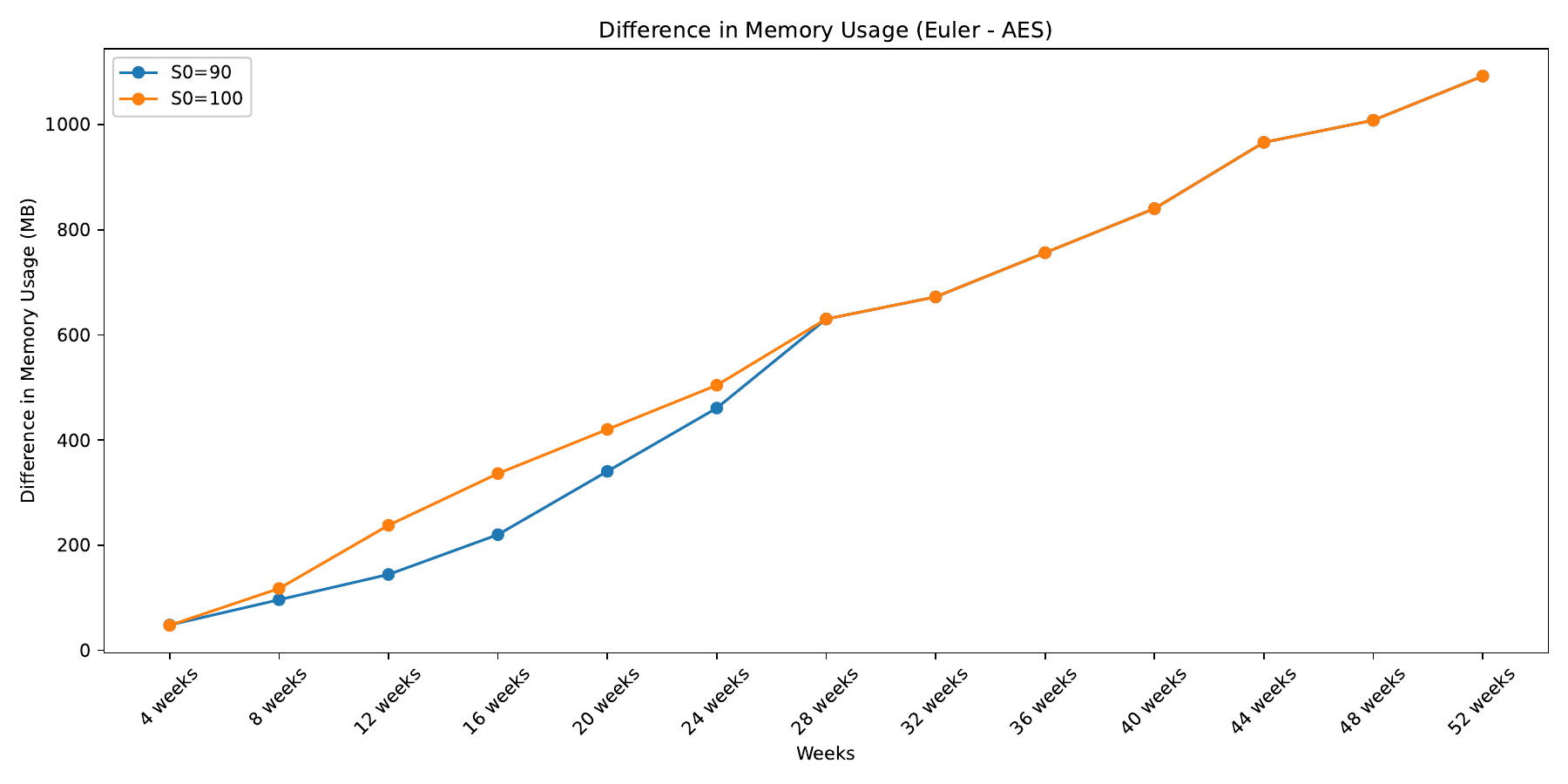}}}
\caption{Comparative analysis of AES and Euler schemes using twice as many steps for Euler scheme simulation: (a) Accuracy difference, (b) Computation time difference, (c) Memory usage difference.}
\label{fig:Difference_AES_Euler_2x}
\end{figure}

\textbf{Bermudan Options Experiment 3.} Building upon the previous experiment, this analysis focuses on in-the-money (\( S_0 = 90 \)) and at-the-money (\( S_0 = 100 \)) options, where the AES scheme consistently demonstrates better accuracy. For out-of-the-money options, the AES scheme does not outperform the Euler scheme, as observed earlier. To further evaluate the accuracy and efficiency trade-offs, the Euler scheme is run with twice as many steps as the AES scheme to provide a fair comparison. The results are summarized in Fig. \ref{fig:Difference_AES_Euler_2x}.

\begin{itemize}
    \item Figure \ref{fig:Difference_AES_Euler_2x}(a): The accuracy difference between the AES and Euler schemes is presented as the Euler scheme's relative error minus the AES's relative error. Positive values indicate the AES scheme achieves lower errors. The results show that the AES scheme maintains an accuracy advantage, particularly for longer times to maturity, where the Euler scheme’s relative error grows more pronounced.
    
    \item Figure \ref{fig:Difference_AES_Euler_2x}(b): The computation time difference highlights the efficiency gains of the AES scheme. While the Euler scheme requires twice as many steps, leading to an increase in computational cost, the AES scheme achieves comparable or better accuracy with fewer steps, making it computationally advantageous, particularly as the option’s time-to-maturity increases.
    
    \item Figure \ref{fig:Difference_AES_Euler_2x}(c): The memory usage difference between the schemes further emphasizes the efficiency of the AES method. As the number of steps doubles for the Euler scheme, memory usage scales significantly. The AES scheme demonstrates clear memory efficiency, especially for longer maturities, where the cost of storing paths and intermediate calculations becomes substantial for the Euler method.
\end{itemize}

These results collectively confirm the advantages of the AES scheme over the Euler scheme. While the Euler scheme performs adequately for short times to maturity, its computational time and memory usage grow disproportionately with the number of steps required to achieve similar accuracy. The AES scheme, by contrast, achieves robust accuracy with minimal steps, making it a more efficient alternative, particularly for in-the-money and at-the-money Bermudan options as the time to maturity increases.

Figure \ref{fig:Difference_AES_Euler_2x} highlights that as the time to maturity increases, the AES scheme becomes increasingly advantageous in terms of both computational time and memory usage. This trend is particularly significant for in-the-money options, where the AES consistently achieves lower relative errors. For at-the-money options, the AES remains competitive, providing comparable accuracy with lower computational cost.

\textbf{American Options Experiment 1.} This experiment investigates the performance of the AES scheme in pricing American put options under the Heston model, specifically focusing on accuracy, computational efficiency, and robustness compared to the Euler scheme.

The first set of parameters, satisfying the Feller condition, is given by Eq. \eqref{eq:ParametersFellerConditionHolds}. The reference prices used to compute the relative errors are obtained as an average of American put option prices reported in the literature.

\begin{table}[htbp]
\centering
\caption{Prices of American options under the Heston model using AES and Euler schemes with minimum ($M=12$) steps.}
\label{table:AmericanOptions}
\begin{tabular}{cccccc}
\toprule
Scheme & $S_0 = 8$ & $S_0 = 9$ & $S_0 = 10$ & $S_0 = 11$ & $S_0 = 12$ \\
\midrule
AES    & 1.9860 & 1.1093 & 0.5190 & 0.2108 & 0.0796 \\
Time (s) & (6.7210) & (6.3420) & (5.3060) & (3.8190) & (3.3500) \\
\midrule
Euler  & 1.9858 & 1.1125 & 0.5235 & 0.2146 & 0.0819 \\
Time (s) & (6.3480) & (5.9440) & (4.8510) & (3.4090) & (2.9420) \\
\midrule
Reference Price & 1.9996 & 1.1079 & 0.5211 & 0.2153 & 0.0833 \\
\bottomrule
\end{tabular}
\end{table}

Given that the time to maturity is $T = 0.25$, according to the previous discussion, the minimum number of steps to price the American option would be $M = 12$. The results in Table \ref{table:AmericanOptions} indicate that both the AES and Euler schemes perform well, producing prices close to the reference values. The AES scheme demonstrates slightly better accuracy for in-the-money (\( S_0 = 8, 9 \)) and at-the-money (\( S_0 = 10 \)) options. However, the Euler scheme is negligibly faster, as reflected in the average computation times.

\textbf{American Options Experiment 2.} This experiment evaluates the performance of the AES scheme under a parameter set where the Feller condition does not hold and the correlation parameter is negative, as defined in Eq. \eqref{eq:ParametersFellerConditionDoesNotHold}. These parameters reflect a more realistic market setting where volatility tends to increase when the asset price declines (negative correlation). The reference prices for this experiment are obtained from \cite{haentjens2015adi}.

\begin{table}[htbp]
\centering
\caption{Prices of American options under the Heston model using AES and Euler schemes, including average computation time in seconds with $M=12$.}
\label{table:Options}
\begin{tabular}{cccc}
\toprule
Scheme & $S_0 = 90$ & $S_0 = 100$ & $S_0 = 110$ \\
\midrule
AES    & 9.9622 & 3.1992 & 0.9078 \\
Time (s) & (6.9850) & (5.2550) & (3.5530) \\
\midrule
Euler  & 9.9646 & 3.2283 & 0.9205 \\
Time (s) & (6.4910) & (4.7400) & (3.0280) \\
\midrule
Reference Price & 9.9784 & 3.2047 & 0.9274 \\
\bottomrule
\end{tabular}
\end{table}

The results in Table \ref{table:Options} show that the AES and Euler schemes produce prices close to the reference values, demonstrating their reliability even under less favorable conditions. The following observations can be made:
\begin{itemize}
    \item For in-the-money options (\( S_0 = 90 \)) and out-of-the-money options (\( S_0 = 110 \)), the Euler scheme achieves slightly better accuracy than the AES scheme. However, the difference in accuracy is minor and comes at the cost of increased computational steps or time for Euler when compared to AES in other scenarios.
    
    \item For at-the-money options (\( S_0 = 100 \)), the AES scheme performs slightly betters. 
\end{itemize}

\begin{figure}
    \centering
    \subfloat[Relative Errors of American put option prices with parameters given in Eq. \eqref{eq:ParametersFellerConditionHolds} calculated using AES.]{%
    \resizebox*{7cm}{!}{\includegraphics{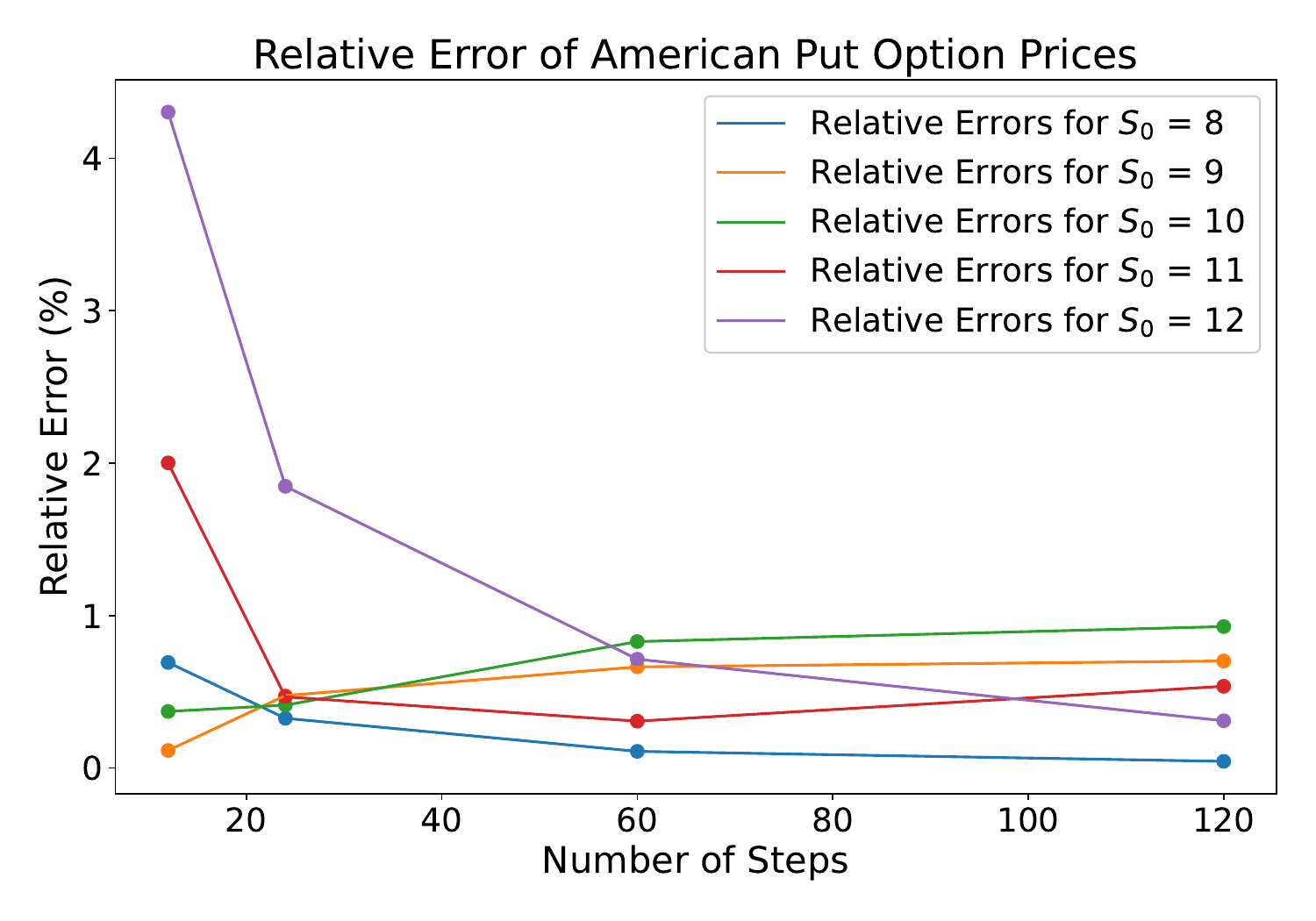}}}\hspace{5pt}
    \subfloat[Relative Errors of American put option prices with parameters given in Eq. \eqref{eq:ParametersFellerConditionDoesNotHold} calculated using AES.]{%
    \resizebox*{7cm}{!}{\includegraphics{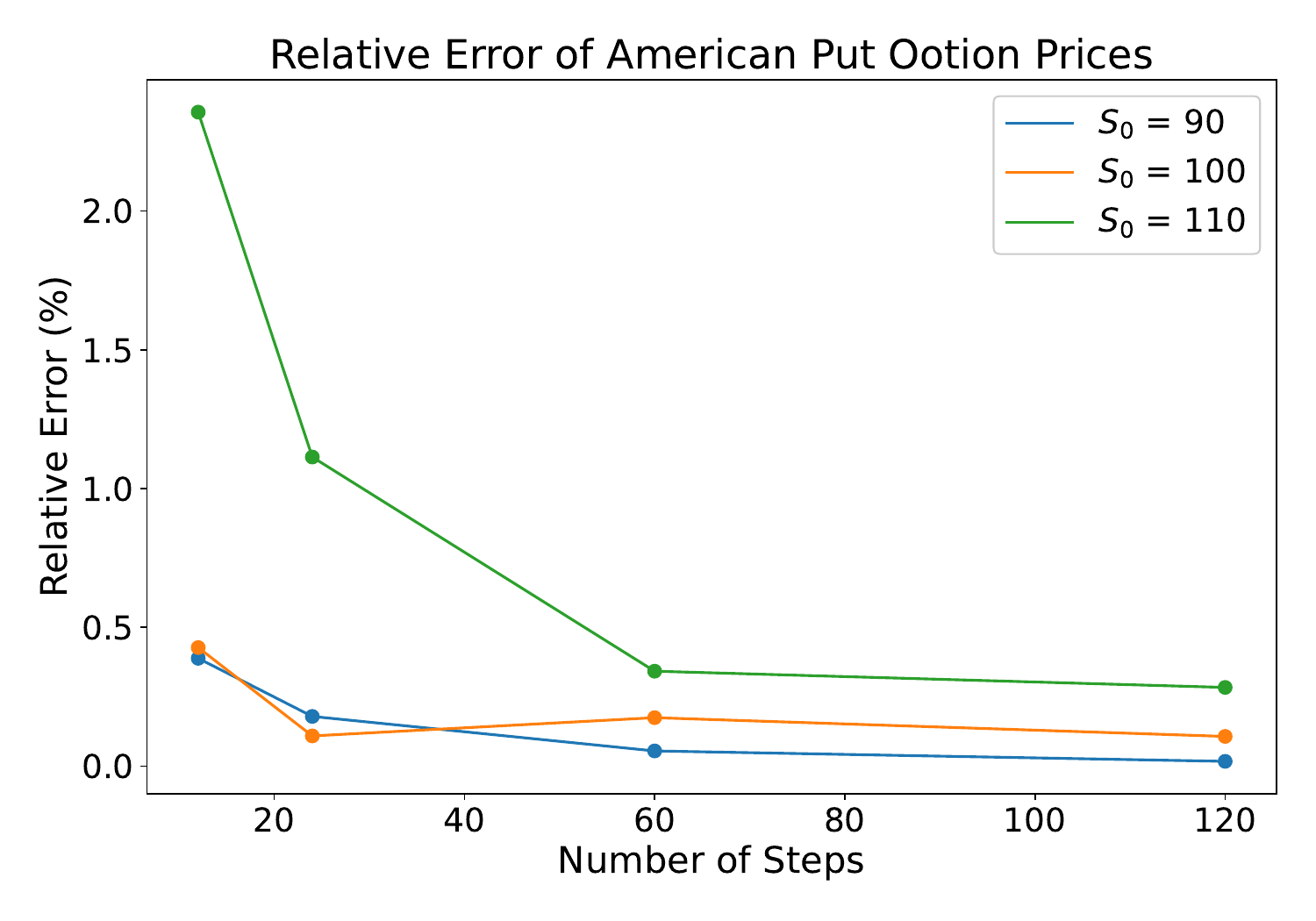}}}
    \caption{Relative Errors of American put option prices calculated using AES under different parameter conditions: (a) Parameters satisfying the Feller Condition, (b) Parameters not satisfying the Feller Condition. The AES scheme achieves high accuracy with a small number of steps.}
    \label{fig:RelativeErrorsCombined}
\end{figure}

To further investigate the behavior of the AES scheme under these conditions, Fig. \ref{fig:RelativeErrorsCombined} presents the relative errors for both parameter sets, highlighting the performance of AES with a minimal number of steps.

\begin{itemize}
    \item Figure \ref{fig:RelativeErrorsCombined}(a): When the Feller condition holds, the AES scheme achieves consistently low relative errors across all asset price levels, even with a small number of steps (\( M = 24 \)). This demonstrates that AES can deliver accurate results without requiring fine time discretization, making it computationally efficient.

    \item Figure \ref{fig:RelativeErrorsCombined}(b): For parameters that violate the Feller condition, the AES scheme still performs well with a small number of steps, particularly for in-the-money and at-the-money options.
\end{itemize}

These findings emphasize the ability of the AES scheme to maintain high accuracy with a minimal number of steps. The AES scheme’s performance highlights its suitability for scenarios where reducing computational cost without sacrificing accuracy is essential.

\textbf{Alternative Perspective on Bermudan Option Pricing.} An alternative perspective is considered to address the limited availability of reference prices for Bermudan options. The AES scheme's capability in pricing Bermudan options has been demonstrated when the number of exercise dates matches the number of time steps. Notably, the pricing methodology for American put options can also serve to approximate Bermudan option prices when the number of exercise opportunities corresponds to the number of simulation steps.

Setting the number of exercise dates to three with $T = 0.25$ (three months) corresponds to pricing Bermudan options with monthly exercise opportunities. Similarly, when $M = 6$, indicating six exercise dates, the resulting price represents the Bermudan option price with bi-monthly exercise opportunities (every two months). This pattern continues for higher values of $M$, indicating different frequencies of exercise dates for the Bermudan options.

Examining Table \ref{table:AmericanAndBermudanTable}, the analysis begins with $M = 3$ steps. This implies that the price obtained for American put options corresponds to the Bermudan option price when the number of exercise dates is 3. As expected, when pricing Bermudan options, the price is generally lower than that of American options due to fewer exercise opportunities. Consequently, the error would be higher when compared to American option prices. However, the error remains acceptable for in-the-money and at-the-money options, as well as for out-of-the-money options when the number of exercise dates is higher.

\begin{table}[htbp]
\centering
\caption{Prices of American put options under the Heston model using AES with parameters given by Eq.~\eqref{eq:ParametersFellerConditionHolds}.}
\label{table:AmericanAndBermudanTable}
\begin{tabular}{cccccc}
\toprule
Number of Steps & $S_0=8$ & $S_0=9$ & $S_0=10$ & $S_0=11$ & $S_0=12$ \\
\midrule
$M=\phantom{10}3$   & 1.942 & 1.079 & 0.491 & 0.190 & 0.067 \\
$M=\phantom{10}6$   & 1.971 & 1.101 & 0.510 & 0.204 & 0.076 \\
$M=\phantom{1}12$   & 1.986 & 1.109 & 0.519 & 0.211 & 0.080 \\
$M=\phantom{1}24$   & 1.993 & 1.113 & 0.523 & 0.214 & 0.082 \\
$M=\phantom{1}60$   & 1.997 & 1.115 & 0.525 & 0.216 & 0.083 \\
$M=120$             & 1.999 & 1.116 & 0.526 & 0.216 & 0.083 \\
\midrule
Reference prices    & 1.999 & 1.108 & 0.521 & 0.215 & 0.083 \\
\bottomrule
\end{tabular}
\end{table}

\subsection{AES under the Double Heston Model}

Attention is now shifted to the AES under the double Heston Model. The objective of this study is to examine the influence of AES on the pricing of American options under the double Heston model when the number of time steps is small and compare it to the Euler scheme. The Bermudan options are not considered due to a lack of reference prices, as there are no decent approximations.

Throughout the numerical analysis, the simulation will consistently have a fixed number of paths, with $N = 1,000,000$, and a total of 20 simulations will be conducted.

Algorithm \ref{alg:DoubleHestonAES} in Appendix A illustrates the implementation of the AES scheme to the Double Heston.

\textbf{American Options.} The set of parameters used for the numerical analysis of the double Heston model and the corresponding reference prices of American put options are adopted from \cite{zhang2019american} and are given by
\begin{equation*}
\begin{aligned}
S_0 &= 61.9, &\quad T &= 0.25, &\quad r &= 0.03, &\quad v^1_0 &= 0.2, &\quad v^2_0 &= 0.49, \\
\gamma_{\nu_{1}} &= 0.1, &\quad \gamma_{\nu_{2}}&= 0.2, &\quad \bar{\nu}_1 &= 0.1, &\quad \bar{\nu}_2 &= 0.15, \\
\kappa_1 &= 0.9, &\quad \kappa_2 &= 1.2, &\quad \rho_{1,3} &= -0.5, &\quad\rho_{2,4} &= -0.5.
\end{aligned}
\label{eq:DoubleHestonModelParameters}
\end{equation*}
The reference prices are obtained using the Modified Asymptotic Expansion (MAE) algorithm. Note that the Feller condition does not hold for this choice of parameters.

Table \ref{table:DoubleHeston} shows that the AES provides significantly higher accuracy than the Euler scheme for in-the-money, at-the-money, and out-of-the-money options when the number of steps is $M = 12$  for a time to maturity of a quarter. Additionally, the AES is more efficient under these conditions.

\begin{table}[htbp]
\centering
\caption{Prices of American options under the double Heston model using AES and Euler schemes with $M=12$ steps, including average computation time in seconds.}
\label{table:DoubleHeston}
\begin{tabular}{cccc}
\toprule
Scheme & $K = 56.9$ & $K = 61.9$ & $K = 66.9$ \\
\midrule
AES    & 6.992 & 9.635 & 12.676 \\
Time (s)  & (134.35) & (156.29) & (170.68) \\
\midrule
Euler  & 5.116 & 7.611 & 10.608 \\
Time (s)  & (165.55) & (191.68) & (207.35) \\
\midrule
Reference Price & 6.887 & 9.504 & 12.520 \\
\bottomrule
\end{tabular}
\end{table}

The next analysis begins by evaluating the prices of American put options using the given parameters. The corresponding prices for varying numbers of steps are shown in Table \ref{table:DoubleHestonAmericanBermudan}.

\begin{table}[htbp]
\centering
\caption{Prices of American put options under the double Heston model using AES.}
\label{table:DoubleHestonAmericanBermudan}
\begin{tabular}{cccc}
\toprule
Number of Steps & $K=56.9$ & $K=61.9$ & $K=66.9$ \\
\midrule
$M=\phantom{1}12$   & 6.995 & 9.636 & 12.678 \\
$M=\phantom{1}24$   & 6.951 & 9.579 & 12.614 \\
$M=\phantom{1}60$   & 6.918 & 9.543 & 12.568 \\
$M=120$             & 6.906 & 9.526 & 12.546 \\
\midrule
Reference prices   & 6.887 & 9.504 & 12.520 \\
\bottomrule
\end{tabular}
\end{table}

The accuracy of the AES in pricing American put options under the double Heston model is high, even with a small number of steps. A satisfactory price can be achieved with as few as $M=12$ steps, which is considered minimal. This observation is evident in Table \ref{table:DoubleHestonAmericanBermudan}, showing that the AES works well for in-the-money, at-the-money, and out-of-the-money options. 

A similar conclusion can be drawn when selecting alternative parameter sets from \cite{zhang2019american}. However, for the sake of convenience, the corresponding results are not presented in this article.

\textbf{Alternative Perspective on Bermudan Option Pricing.} A similar pattern observed with the AES under the Heston model can be discussed here. Specifically, the American option prices obtained can be considered as Bermudan option prices when the number of exercise dates equals to the number of steps used for simulation. Since it has been demonstrated that the relative errors remain relatively small during the pricing of American options, an increase in these errors is anticipated when considering American option prices as Bermudan option prices. Nevertheless, even accounting for this, the prices remain reasonable when the number of exercise dates matches the number of steps.

\section{Conclusion and Future Work}\label{Sec:Conclusion}
In conclusion, this paper has introduced an AES scheme for pricing Bermudan and American options under Heston-type stochastic volatility models, extending it analytically to the double Heston model. The AES method enhances simulation accuracy compared to traditional numerical methods, such as the Euler-Maruyama scheme, particularly when the number of time steps is minimal for in-the-money and at-the-money options.

The AES scheme demonstrates several strengths. By achieving high accuracy with minimal time steps, it reduces computational time and memory usage, making it efficient for simulations that require fine time discretization. Furthermore, the use of the noncentral chi-square distribution ensures the positivity of the variance process, addressing a key limitation of traditional discretization methods where various truncation techniques are often necessary. These features establish the AES scheme as a robust and computationally efficient approach for pricing options in stochastic volatility models.

However, the study is not without its limitations. Although the AES scheme performs exceptionally well for in-the-money and at-the-money options, its accuracy for out-of-the-money options with fewer exercise opportunities remains slightly lower compared to Euler-based methods. Additionally, the scope of this study is limited to the Heston and double Heston models, leaving the generalization of the AES scheme to other stochastic volatility models, such as those with jumps or higher dimensions, for future work.

Future work will focus on addressing these limitations and expanding the scope of the AES scheme. This includes extending the methodology to other multifactor stochastic volatility models. Additionally, the performance of the AES scheme in high-dimensional settings will be explored, where traditional methods often struggle due to the curse of dimensionality. To further enhance efficiency and accuracy, machine learning techniques will be incorporated, particularly for solving complex option pricing tasks. 

In summary, the AES scheme provides a promising alternative to traditional simulation methods for pricing Bermudan and American options. Its ability to maintain high accuracy and computational efficiency, even with minimal time steps.

\section{Appendices}

\noindent\textbf{Appendix A. Implementation Schemes}\medskip

In the AES schemes, the variance $\nu_{i+1}$ at time $t_{i+1}$ is drawn exactly from the noncentral chi-squared distribution followed by the CIR process, ensuring positive variance values. In contrast, it is well known that under the standard Euler discretization scheme the variance $\nu_{i+1}$ can become negative (see, e.g., \cite{oosterlee2019mathematical}). To handle this problem, the following \emph{truncated Euler scheme} is a common approach (\cite{lordetal}) and is used in all experiments where the Euler scheme is implemented.
\[  \nu_{i+1}= \left(\nu_{i} +\kappa\left(\bar{\nu}-\nu_{i}  \right) \Delta t+\gamma \sqrt{\nu_{i}\Delta t} Z \right)^+, \]
where $Z$ is standard normal random variable. For the double Heston model, each of the two variance processes is simulated in this manner. 

For experiments implementing Euler scheme, at each time step, the variance is simulated using the above truncated Euler scheme, then the asset price process is simulated following the standard Euler scheme.  Details on how the variance and asset processes are simulated using the AES schemes are given in the following algorithms.

\begin{algorithm}[H]
\caption{AES Scheme for Heston Model}
\label{alg:AES}
\begin{algorithmic}[1]

\Require $M$: Number of time steps; $T$: Maturity; $r, S_0, \kappa, \gamma, \rho, \bar{\nu}, v_0$: Model parameters
\Ensure Simulated path $\{S_j, V_j\}$ for $j=0,\dots,M$

\vspace{0.5em}

\State \textbf{Set up the time grid:}
\[
  \Delta t \gets \frac{T}{M}, \quad t_j \gets j \Delta t.
\]

\State \textbf{Initialize:} 
\[
  V_0 \gets v_0, \quad X_0 \gets \ln(S_0).
\]

\State \textbf{Generate Brownian increments.}

\For{$j = 0$ to $M-1$}
   \State \textbf{Update variance:} 
   Sample $V_{j+1}$ from the CIR distribution using Eq. \eqref{eq:HestonAESVariance}:
   \[
     V_{j+1} \gets \text{CIR\_Sample}(\kappa,\gamma,\bar{\nu},\Delta t,V_j).
   \]

   \State \textbf{Compute coefficients for the log--price:}
   \[
     c_0 = \left(r - \frac{\rho \kappa \bar{\nu}}{\gamma}\right)\Delta t,\quad
     c_1 = \left(\frac{\rho \kappa}{\gamma}-0.5\right)\Delta t - \frac{\rho}{\gamma},\quad
     c_2 = \frac{\rho}{\gamma}.
   \]

   \State \textbf{Update log-price:}
   \[
     X_{j+1} \gets X_j 
     + c_0 
     + c_1 V_j 
     + c_2 V_{j+1}
     + \sqrt{1-\rho^2}\,\sqrt{V_j}(W^{(1)}_{j+1}-W^{(1)}_{j}).
   \]
\EndFor

\State \textbf{Recover asset prices:}
\[
  S_j \gets e^{X_j} \text{ for } j=0,\dots,M.
\]

\State \textbf{Return:} $\{S_j, V_j\}$

\end{algorithmic}
\end{algorithm}

\begin{algorithm}[H]
\caption{AES Scheme for the Double Heston Model}
\label{alg:DoubleHestonAES}
\begin{algorithmic}[1]

\Require 
$M$: Number of time steps; 
$T$: Maturity; 
$r, S_0$: Interest rate, initial price;
$\kappa_1, \gamma_{\nu_1}, \bar{\nu}_1, v_{0,1}$, $\kappa_2, \gamma_{\nu_2}, \bar{\nu}_2, v_{0,2}$: Parameters for the variance processes;
$\rho_{1,3}, \rho_{2,4}$: Correlations.

\Ensure Simulated path $\{S_j, \nu_j^{(1)}, \nu_j^{(2)}\}$ for $j=0,\dots,M$.

\vspace{0.5em}

\State \textbf{Set up the time grid:}
\[
  \Delta t \gets \frac{T}{M}, \quad t_j \gets j \Delta t.
\]

\State \textbf{Initialize:} 
\[
  \nu_0^{(1)} \gets v_{0,1}, \quad \nu_0^{(2)} \gets v_{0,2}, \quad X_0 \gets \ln(S_0).
\]

\State \textbf{Precompute constants:}
\[
\begin{aligned}
c_0 &= \biggl(r - \frac{\rho_{1,3}\kappa_1\bar{\nu}_1}{\gamma_{\nu_1}} - \frac{\rho_{2,4}\kappa_2\bar{\nu}_2}{\gamma_{\nu_2}}\biggr)\Delta t, & 
c_3 = \frac{\rho_{1,3}}{\gamma_{\nu_1}}, & 
c_5 = (1 - \rho_{1,3}^2)\Delta t, \\
c_1 &= \biggl(\frac{\rho_{1,3}\kappa_1}{\gamma_{\nu_1}} - \frac{1}{2}\biggr)\Delta t - \frac{\rho_{1,3}}{\gamma_{\nu_1}}, &
c_4 = \frac{\rho_{2,4}}{\gamma_{\nu_2}}, &
c_6 = (1 - \rho_{2,4}^2)\Delta t, \\
c_2 &= \biggl(\frac{\rho_{2,4}\kappa_2}{\gamma_{\nu_2}} - \frac{1}{2}\biggr)\Delta t - \frac{\rho_{2,4}}{\gamma_{\nu_2}}.
\end{aligned}
\]

\State \textbf{Generate Brownian increments.}

\For{$j = 0$ to $M-1$}

   \State \textbf{CIR updates for variances:}
   Compute:
\begin{align*}
     \nu_{j+1}^{(1)} &\gets \bar{c}^{(1)}(t_{j+1},t_j) \cdot \chi^2\bigl(\delta^{(1)}, \bar{\kappa}^{(1)}(t_{j+1},t_j)\bigr), \\
     \nu_{j+1}^{(2)} &\gets \bar{c}^{(2)}(t_{j+1},t_j) \cdot \chi^2\bigl(\delta^{(2)}, \bar{\kappa}^{(2)}(t_{j+1},t_j)\bigr),
\end{align*}
   where for $j=1,2$:
   \[
     \bar{c}^{(j)} = \frac{\gamma_{\nu_j}^2}{4\kappa_j}(1 - e^{-\kappa_j \Delta t}), \quad
     \bar{\kappa}^{(j)} = \frac{4\kappa_j \nu_j^{(j)}e^{-\kappa_j \Delta t}}{\gamma_{\nu_j}^2(1 - e^{-\kappa_j \Delta t})}, \quad
     \delta^{(j)} = \frac{4\kappa_j \bar{\nu}_j}{\gamma_{\nu_j}^2}.
   \]

   \State \textbf{Update log-price:}
\begin{align*}
    X_{j+1} &\gets X_j
     + c_0 
     + c_1 \nu_j^{(1)} 
     + c_2 \nu_j^{(2)} 
     + c_3 \nu_{j+1}^{(1)} 
     + c_4 \nu_{j+1}^{(2)} \\
     &\phantom{\gets}+ \sqrt{c_5 \nu_j^{(1)}}(W_{j+1}^{(1)} - W_{j}^{(1)})
     + \sqrt{c_6 \nu_j^{(2)}}(W_{j+1}^{(2)} - W_{j}^{(2)}).
\end{align*}

\EndFor

\State \textbf{Recover asset prices:}
\[
  S_j \gets \exp(X_j) \quad \text{for } j=0,\dots,M.
\]

\State \textbf{Return:} $\{S_j, \nu_j^{(1)}, \nu_j^{(2)}\}$.

\end{algorithmic}
\end{algorithm}

\appendix

\end{document}